\documentclass[aps, prb, twocolumn, amsmath, amssymb]{revtex4}

\usepackage{graphicx}  
\usepackage[utf8]{inputenc} 
\usepackage{url}
\usepackage{dcolumn}	
\usepackage{bm} 
\usepackage{mathtools}
\usepackage{float}
\usepackage{placeins}
\usepackage{physics}
\usepackage{subfig}
\usepackage{caption}
\usepackage{hyperref}
\begin{document}

\title{Investigating the relation between chaos and the three body problem} 

\author{T.S.Sachin Venkatesh \\ \href{mailto:tssachin.venkatesh@gmail.com}{\nolinkurl{tssachin.venkatesh@gmail.com}}}
\affiliation{Delhi Technological University, 
Delhi 110042, India}
\author{Vishak Vikranth}
\affiliation{Bengaluru 560076, India} 

\begin{abstract}
    We review the properties of fractals, the Mandelbrot set and how deterministic chaos ties to the picture. A detailed study on three body systems, one of the major applications of chaos theory was undertaken. Systems belonging to different families produced till date were studied and their properties were analysed. We then segregated them into three classes according to their properties. We suggest that such reviews be carried out in regular intervals of time as there are an infinite number of solutions for three body systems and some of them may prove to be useful in various domains apart from hierarchical systems.
    
    Key words - Celestial mechanics, Three-body problem, Gravitational interaction, Chaos, Orbits, Astronomical simulations
    
\end{abstract}
\maketitle
\section{\label{sec:intro}Introduction}
Exploring the connections between different theories of mathematics leads one through successive topics which diverge very little from one another, but looking at the the path as a whole, the initial point of probing and the final point have very little in common. So a quick introduction of the topics we covered is listed below

\subsection{Fractals}
Fractals are infinitely complex patterns that are self-similar across different scales. They are created by repeating a simple process over and over in an ongoing feedback loop. A key characteristic of a fractal is its fractal dimension. Unlike  the Euclidean dimension, fractal dimension is generally expressed by a non-integer and is an indicator of the complexity or roughness of a given figure. Some common examples include Sierpiński triangle having a Hausdorff dimension of 1.585, Koch's snowflake with a Hausdorff dimension of 1.262 and the coastline of Britain, whose fractal dimension is 1.21.

\subsection{The Mandelbrot set and the logistic map}
One of the most interesting fractals, the Mandelbrot set is the set of complex numbers c for which the function ${x_{n+1} = x_n^2 + c}$ does not diverge to infinity when iterated from z=0. The path of all such orbits when plotted give us an image which is intuitive and rather easy to understand. All the points inside the main cardioid have a single fixed point, all the points inside the main bulb have 2 limit points and subsequently all other bulbs represent the set of numbers which have different number of limit points. 
The logistic map is a polynomial mapping of degree 2, often cited as an archetypal example of how complex, chaotic behaviour can arise from very simple non-linear dynamical equations. By simple algebraic manipulation, the logistic map ${x_{n+1} = rx_n(1 - x_n)}$, can be recoded into the form ${x_{n+1} = x_n^2 + c}$. This leads us to further explore a relation which produces a rather interesting relation when we change our Point of View. From the branching structure of the logistic bifurcation diagram we can read the cycle number of the corresponding features of the Mandelbrot set.
\begin{figure}
    \centering
    \includegraphics[scale=0.18]{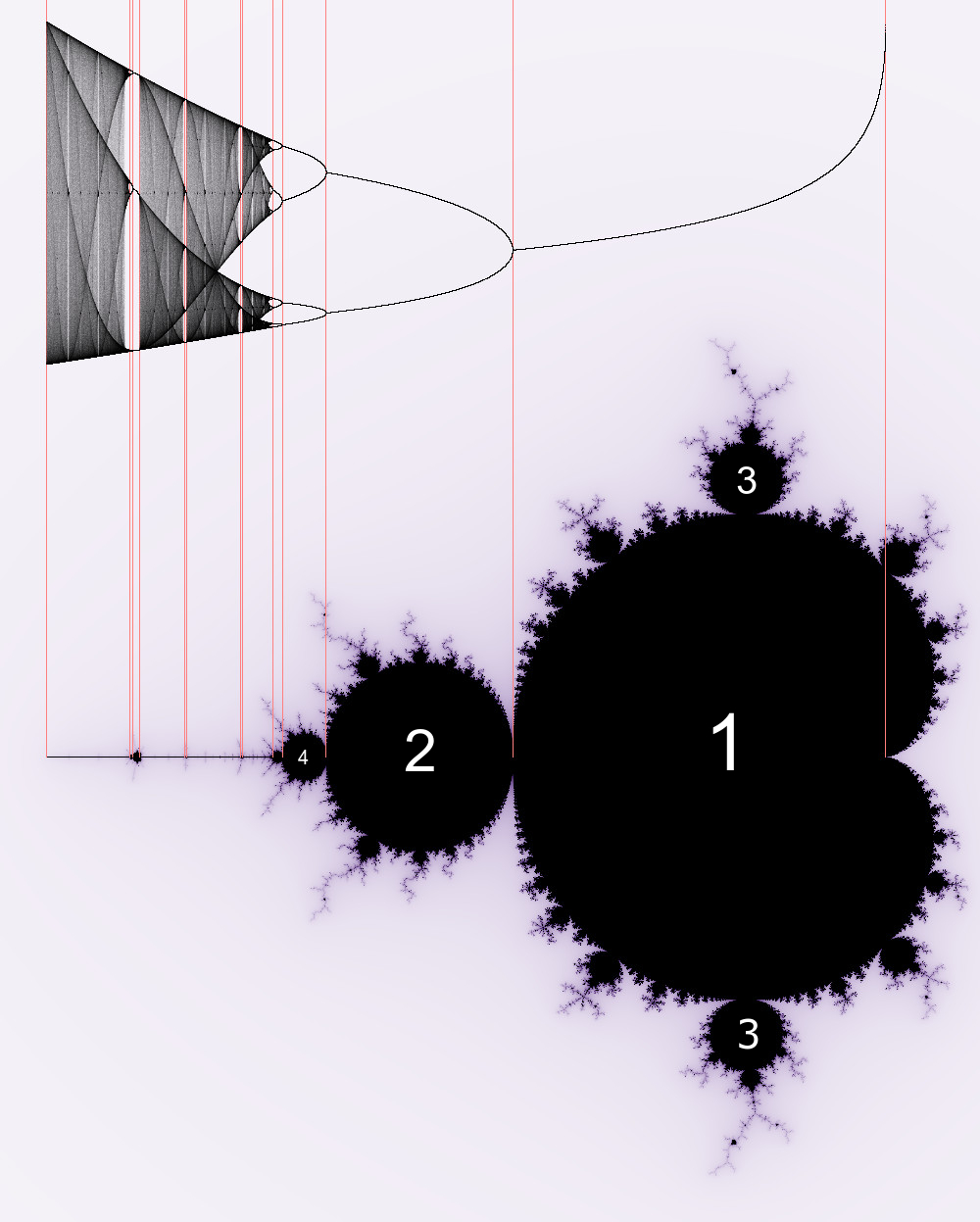}
    \caption{The real line on the mandelbrot set lines up with the bifurcations in the logistic map \cite{Relation}}
\end{figure}

\subsection{Deterministic Chaos}
Deterministic chaos is the study of how systems that follow simple, straightforward, deterministic laws can exhibit very complicated and seemingly random long term behavior. One of the foundations of chaos theory is Edward Lorenz's discovery of systems where the dynamics are sensitive to minute changes in the initial conditions thus are unpredictable. Due to this sensitivity the behaviour of systems appear to be random although model is deterministic, which means that it is well defined and it does not contain any random parameters. In the beginning, this theory was used only in the field of meteorology but soon it was realized that it can also define the other chaotic systems in varied science disciplines which were based on the predictability of future behavior of the systems. 

\section{Theory}
\subsection{The relation between Logistic Map and Chaos}
The logistic equation is of the form ${x_{n+1} = rx_n(1 - x_n)}$. By changing the values of $r$, the following behaviour is observed : When  $r$ is in $[0,1]$, the sequence will converge to 0. When  $r$ is in $[1,3)$, the sequence approaches the value $r - \frac{1}{r}$, regardless of the value for ${x_0}$;though the rate of convergence of the function decreases as the value of $r$ increases. When $r$ is in $[3,1+\sqrt{6})$,the function oscillates between two limit points whose values are determined by $r$. When $r$ is in $[1+\sqrt{6},\sim(3.54409)]$,the sequence will approach permanent oscillations among four limit points. When $r$ increases beyond $\sim(3.54409)$,the function will approach oscillations among eight limit points, then $16$, $32$,and so on. The ratio between the lengths of two successive bifurcations approaches the Feigenbaum constant $\delta$$\approx4.66920$\cite{feigenbaum_constant}. At $r$ $\approx3.56995$ is the onset of chaos. We no longer see oscillations of finite period. Slight variations in the initial ${x_0}$ value yield drastically different results over time.

For values of $r$ beyond $\sim3.56995$ the sequence exhibits chaotic behaviour, but there are still certain values of $r$ that show non-chaotic behavior; these are sometimes called islands of stability. For example for r=$1+\sqrt{8}$ the sequence oscillates between three values, and for slightly higher values of $r$ oscillation among $6$ limit points, then 12 etc.

\subsection{Noise versus Deterministic Chaos}
Noise is the random variation of values whereas chaos happens when the initial state variables of the system differ ever so slightly which lead to drastically different outcomes making it impossible to predict the initial value from the output.

We can also differentiate between noise and deterministic chaos using fractals. The Hausdorff dimension of the attractor of a random model is usually infinite. But the attractor of a model of deterministic chaos always gives us a non integral, finite value. Hence, if the Hausdorff dimension of a model is finite it is good indicator that it is a deterministic model that can be represented by a system of non linear differential equations\cite{dimension_of_attractors}.

\subsection{Lorenz Attractor}
An attractor is a point, set or even a fractal (strange attractor) around which it the function seems to converge. Edward Lorenz, an American physicist and meteorologist who pioneered chaos theory is famous for demonstrating chaotic motion using "Lorenz water wheel": a waterwheel with holes at the bottom and a constant stream pouring from above. The emptying and refilling of the wheel produces unpredictable changes in its direction of rotation and angular velocity.

Lorenz's paper in 1963,tries to model weather and used these non linear differential equations \cite{Lorenz2004}:

\begin{eqnarray}
    \dv{x}{t}=\sigma(y-x) \nonumber \\
    \dv{y}{t}=x(\rho-z)-y \nonumber\\
    \dv{z}{t}=xy-\beta z \nonumber \\
\end{eqnarray}

According to Lorenz's account, while working on the weather model he was running simulations and wanted to repeat a previous simulation. He started the simulation from a random iteration and since it was a computer program it should have given him the exact same results. But the results were in stark disagreement. He initially shrugged this off as an error, but then realized he had not entered the initial conditions exactly. The computer was taking values precise upto six decimal places but the printer only displayed three. He re-entered the rounded off values and this minuscule error caused drastic changes in the outcome.

\begin{figure}[ht]
    \centering
    \includegraphics[scale=0.4]{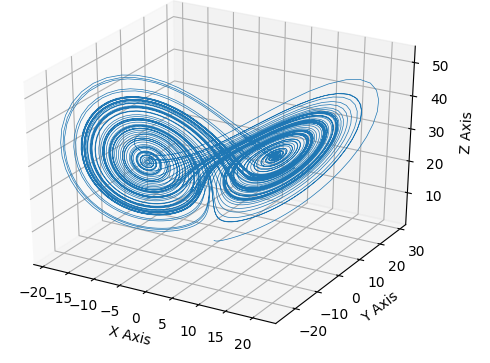}
    \caption{Lorenz attractor\cite{Attractor_image}}
\end{figure}
\FloatBarrier

Lorenz considered the case where $\sigma$ = 10, $\beta$ = 8/3, $\rho$ = 28 with $({x_0}, {y_0}, {z_0}) = (0, 1, 0)$. This gives us the Lorenz attractor\cite{Lorenz2004}.

\subsection{Three Body Problem}
The three body problem is a system of ordinary differential equations modelling three bodies of masses under mutual gravitation in two or three dimensions. The motion of such a system was first pondered upon by Newton and was reconsidered by many mathematicians and scientists. The famous three-body problem has had a great influence on physics, mathematics and non-linear dynamics. It has paved way to a new field in modern science, chaotic dynamics. 

Until 1975, there were not many models of three body systems and the ones that existed were too restrictive and specific about the parameters of the model. In 1890, Poincaré proved the non existence of the uniform first integral of a three-body problem in general, and also highlighted the sensitive dependence to initial conditions of its trajectories\cite{poinc}. Three body systems without mass hierarchy are never thought to be stable for very long\cite{hierarchy}. They can certainly exist for some period of time, but they aren't found to be long term stable. In these systems, each body orbits the center of mass of the system. Mostly, two of the bodies form a close binary system, and the third body orbits this binary at a distance much larger than that of the orbit of the close binary. This arrangement is called hierarchical. The reason for this behaviour is that if the inner and outer orbits are comparable in size, the system may become dynamically unstable, leading to a body being ejected from the system. And if the system in question consists of disproportionate mass bodies, the Hill sphere\cite{hillsp} mechanism comes into play. 

A lot of work has been carried out and various models have been explored such as Poincaré's planar circular restricted three body problem (PCR3BP)\cite{poinc}, Sitnikov models in the early 1900s and the recent discovery of thirteen families of stable planar equal mass three body systems. The recent discoveries have formulated a new method to check for duplicate orbits and sort different instances, they use an abstract space called a `shape space sphere'\cite{yarnarxiv} which describes the shape of the orbits in terms of the relative distances between the objects. Three spots around the sphere's equator mark where two of the particles would collide, and a line drawn over the ball, which must avoid those spots, maps how near the objects get to each other. One of their original solutions nicknamed `yarn' is shown.

\begin{figure}
    \centering
    \begin{minipage}{0.45\textwidth}
        \centering
        \includegraphics[width=0.9\textwidth]{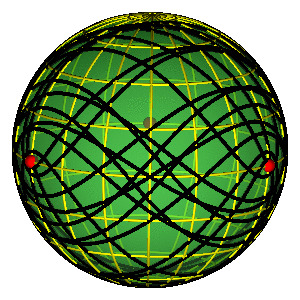} 
        \caption{Shape sphere showing the relative positions of the orbits in a 3-body problem.\cite{yarnweb}}
    \end{minipage}\hfill
    \begin{minipage}{0.45\textwidth}
        \centering
        \includegraphics[width=0.9\textwidth]{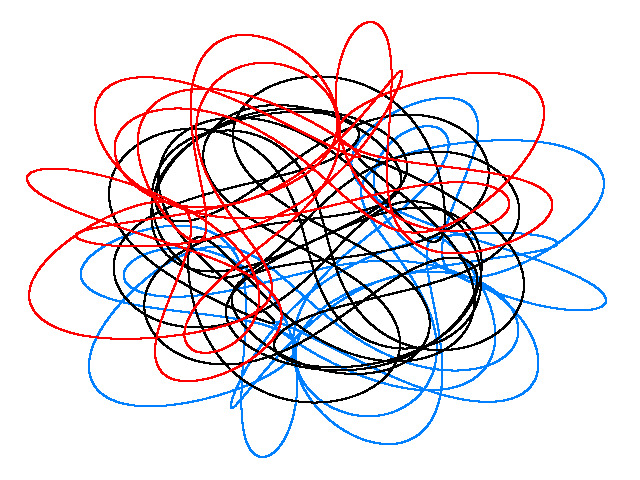} 
        \caption{The same orbits in real euclidean space relative to centre of mass of the system.\cite{yarnweb}}
    \end{minipage}
\end{figure}

\section{Modeling and Observations}
We simulated different models of the three body system ranging from the more popular `Montgomery 8'\cite{mont8} orbit to custom made orbits like `perturbed circular orbits'. Most of these models are equal mass systems, but there are some special orbits that were simulated for unequal mass systems, which were observed to remain stable over extensive number of iterations. The simulated models can broadly be classified into three classes of stability; namely, Long-term stable, Quasi-stable and Chaotic systems.

\subsection{Long term stable}
\begin{figure*}[htp]
    \centering
    \begin{minipage}{0.45\textwidth}
        \centering
        \includegraphics[width=0.9\textwidth]{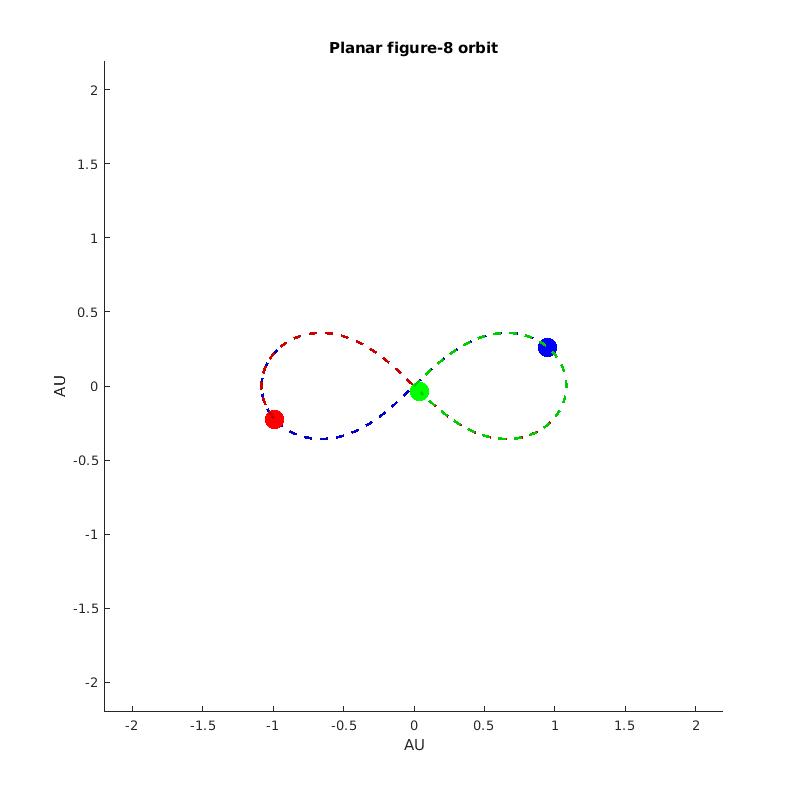} 
        \caption{Montgomery 8 - the system comprises of 3 equal masses with a net angular momentum of zero.}
    \end{minipage}\hfill
    \begin{minipage}{0.45\textwidth}
        \centering
        \includegraphics[width=0.9\textwidth]{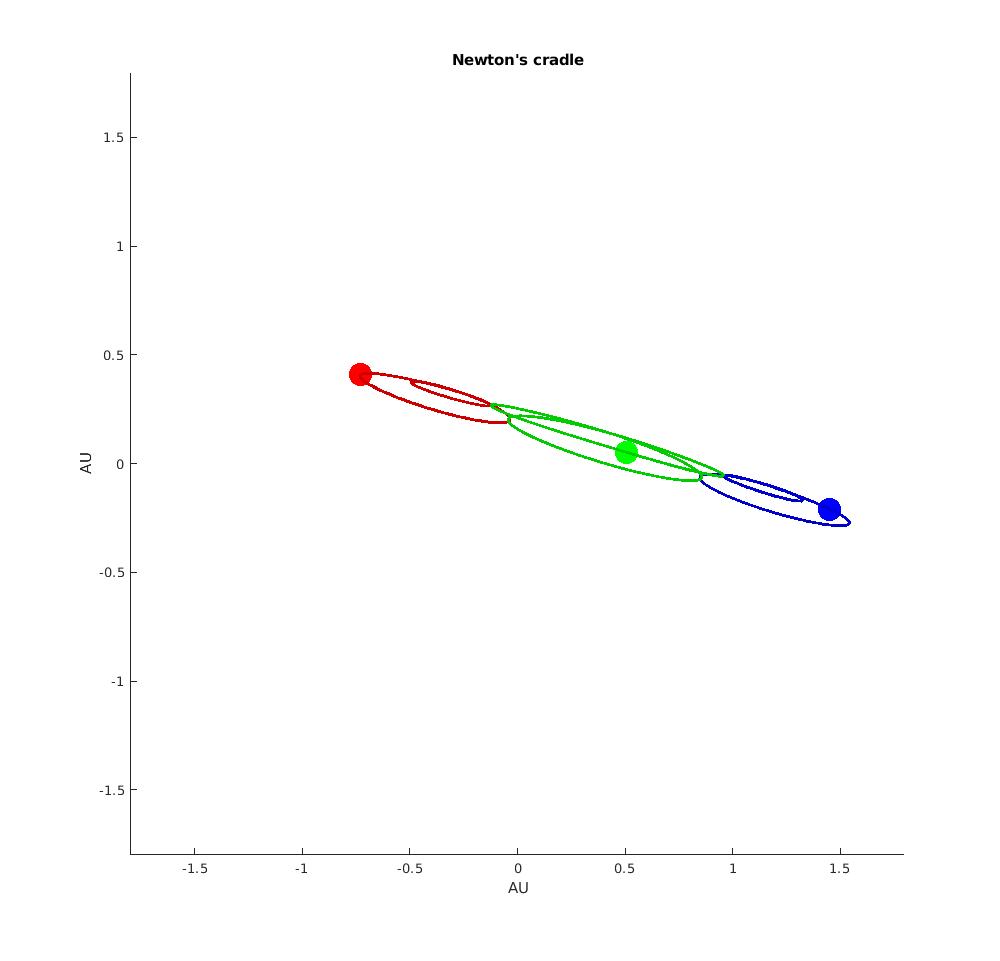} 
        \caption{Newton's cradle-esque system of equal masses that meet at only two points.}
    \end{minipage}
    \begin{minipage}{0.45\textwidth}
        \centering
        \includegraphics[width=0.9\textwidth]{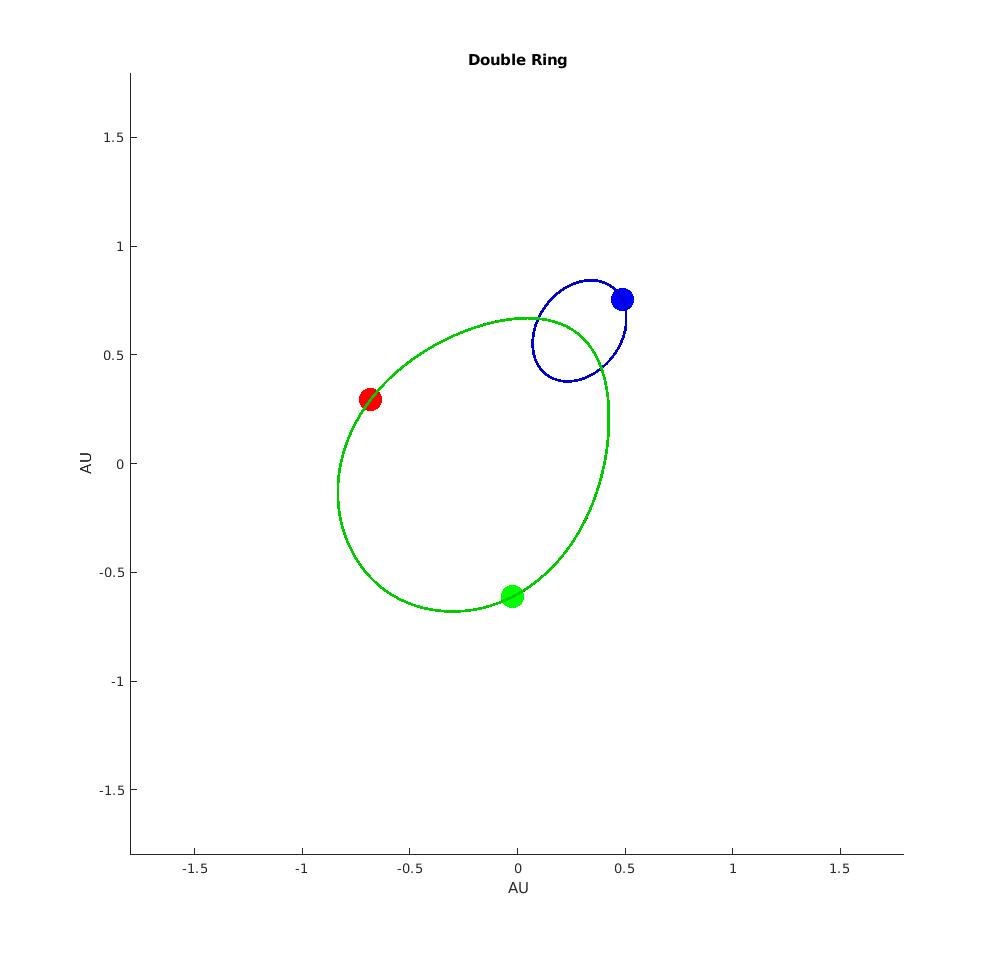} 
        \caption{Double ring system. The blue body acts like a field regulator.}
    \end{minipage}
    \captionsetup{labelformat=empty}
\end{figure*}

The families of such three body systems are extremely rare to compute in a general setup due to the increased complexity in computation owing to the increase in the number of parameters. Some of the first solutions to the three body system were given by Euler and Lagrange for special cases like `Circular Restricted 3 Body Problem' and `Planar Restricted 3 Body Problem'. Until 2013, specific solutions could be sorted into just three families: the Lagrange-Euler family, the Broucke-Hénon family, and the figure-eight family. Since then there have been more than a thousand periodic solutions for the planar three body problem and 13 families of solutions to the more general, non planar three body problem.

The Montgomery 8 orbit uses a variational method to exhibit a simple periodic orbit for the Newtonian problem of three equal masses in the plane\cite{mont8}. 

Other stable orbits include Newton's cradle-esque orbit (FIG. 6) which looks as if two binary orbits have coalesced together to form a system with a period of two. There are two separate levels of orbital systems intertwined into one. The bodies try to slingshot each other but owing to their gravitational pull, they aren't able to escape the orbit, thus when they try to slingshot, they form a larger orbit but when they are unable to exit the system, they are dominated by the gravitational potential and form a shorter orbit.

A more rational system that has a higher probability of being detected in the universe is the system shown below (FIG. 7). This is a solution for an equal mass system where two bodies move in the same orbit while the third body acts as the field regulator, making sure the other two bodies stay in the same orbit. The probability of the existence of such system is relatively high due to the simplicity of the system and the lack of complex orbits which are very sensitive to even the slightest of perturbations.

\subsection{Quasi-stable}
\begin{figure*}[htp]
    \centering
    \begin{minipage}{0.45\textwidth}
        \centering
        \includegraphics[width=1\textwidth]{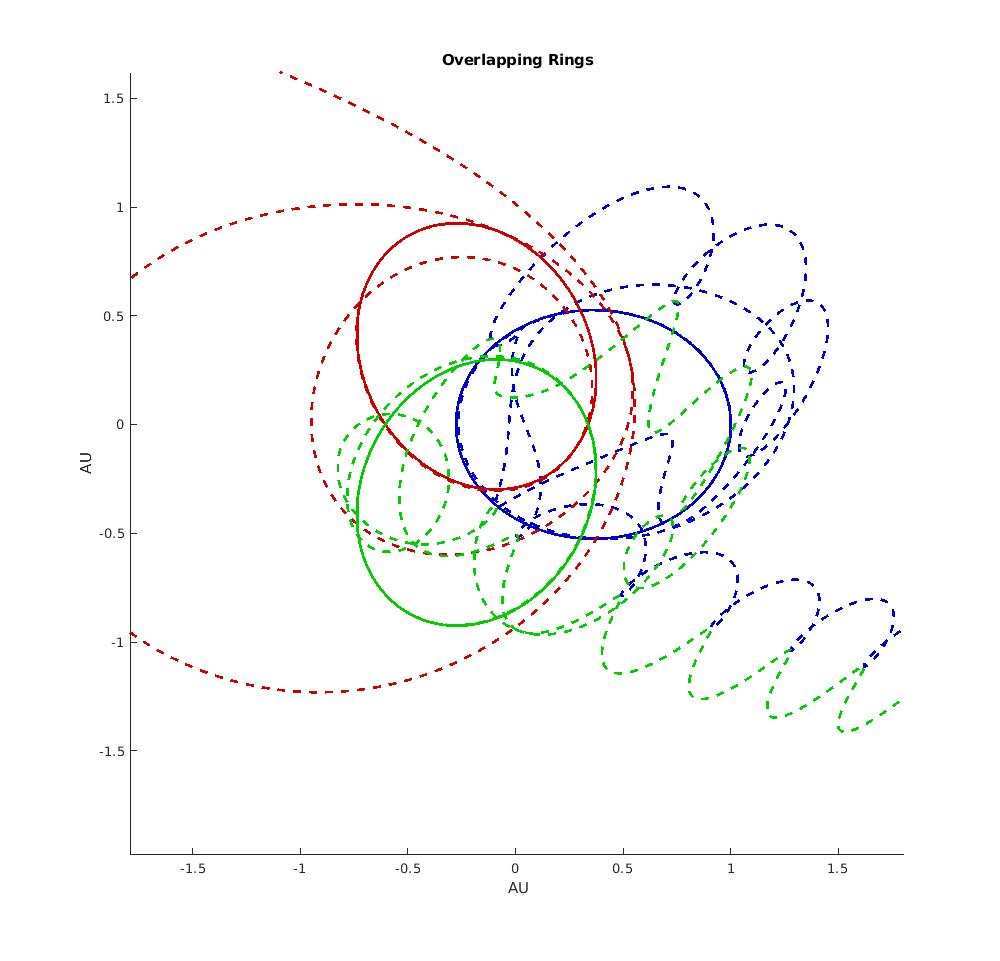} 
        \caption{Overlapping rings with a common centroid. The equal mass bodies are placed at the vertices of an equilateral triangle. The highlighted curves represent the quasi-stable orbits of the systems, while the dashed lines represent the path of the bodies after destabilizing.}
    \end{minipage}\hfill
    \begin{minipage}{0.45\textwidth}
        \centering
        \includegraphics[width=1\textwidth]{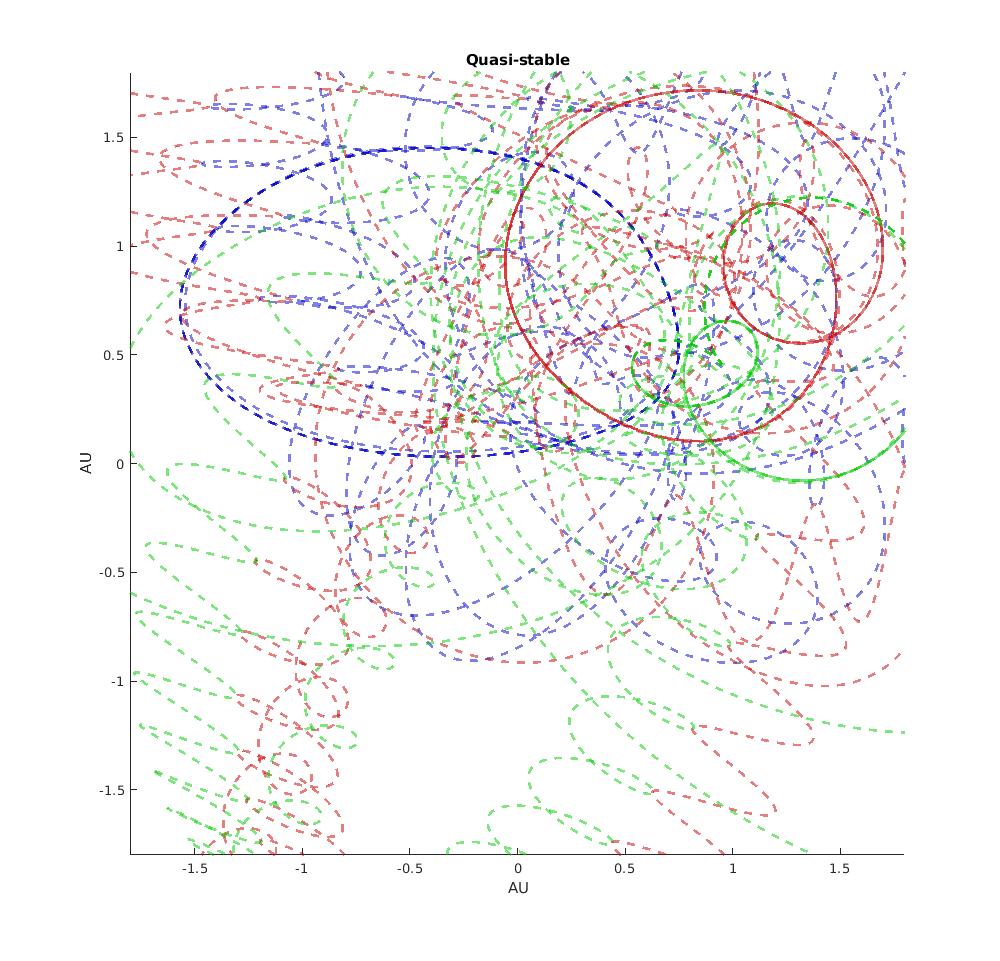} 
        \caption{Unequal mass system destabilised due to perturbations. The highlighted curves represent the quasi-stable orbits of the systems, while the dashed lines represent the path of the bodies after destabilizing.}
    \end{minipage}
\end{figure*}

Three body systems corresponding to this family are much harder to find since these systems remain stable for some period of time which may range anywhere from a few moments to a few hundreds of years but over longer periods of time, they seem to diverge from their orbits and the three body system breaks down to a two body system with the third body being flung away. Such systems are perfect examples to demonstrate sensitive dependence to initial conditions (SDIC), even the slightest changes in the initial conditions of positions, velocity, mass can alter their course and send them travelling in a completely different direction or configuration. 

One such system is where the equal mass bodies are initially placed at the vertices of an equilateral triangle and the centre of mass of the system lies at it's centroid. The bodies revolve around the centre of mass in a symmetric orbit but after some period of time they are flung away from their seemingly stable orbit and chaotic motion takes over the system. Changing the initial conditions such that the bodies now lie at the vertices of a slightly bigger equilateral triangle produces the same type of orbit but the chaotic regime of this configuration is completely different from that of the initial configuration.

An unequal mass system that demonstrates similar behaviour is shown alongside (FIG. 9). The blue body has double the mass of other bodies. When the bodies approach the centre of mass of the system (which is slightly shifted towards blue), the other bodies go around in loops. This perturbation destabilises the system.

\setcounter{figure}{9}
\subsection{Chaotic systems}
\begin{figure*}[htp]
    \centering
    \begin{minipage}{0.45\textwidth}
        \centering
        \includegraphics[width=0.9\textwidth]{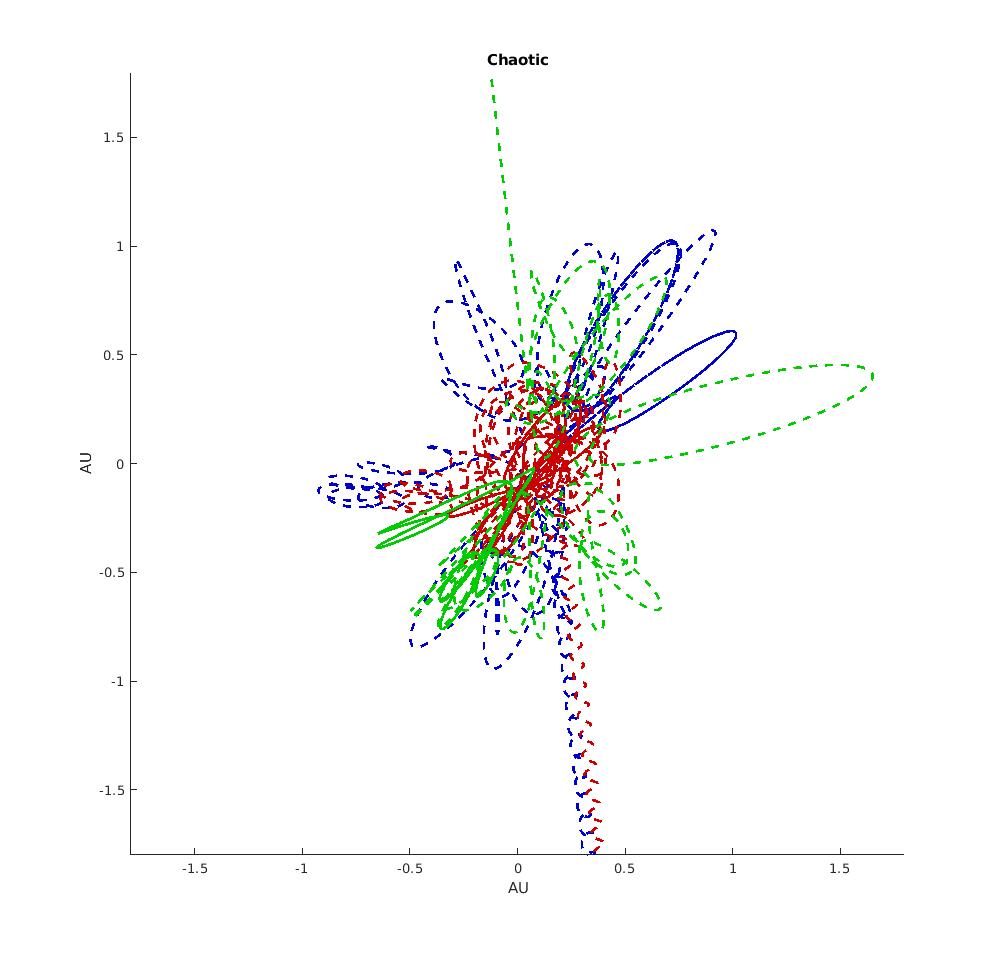} 
        \caption{Chaotic behaviour of an equal mass system. The highlighting feature displayed in both systems is how the three body system reduces to a two body system to attain stability. The third body is always flung away with the help of the gravity assist of other two bodies.}
    \end{minipage}\hfill
    \begin{minipage}{0.45\textwidth}
        \centering
        \includegraphics[width=0.9\textwidth]{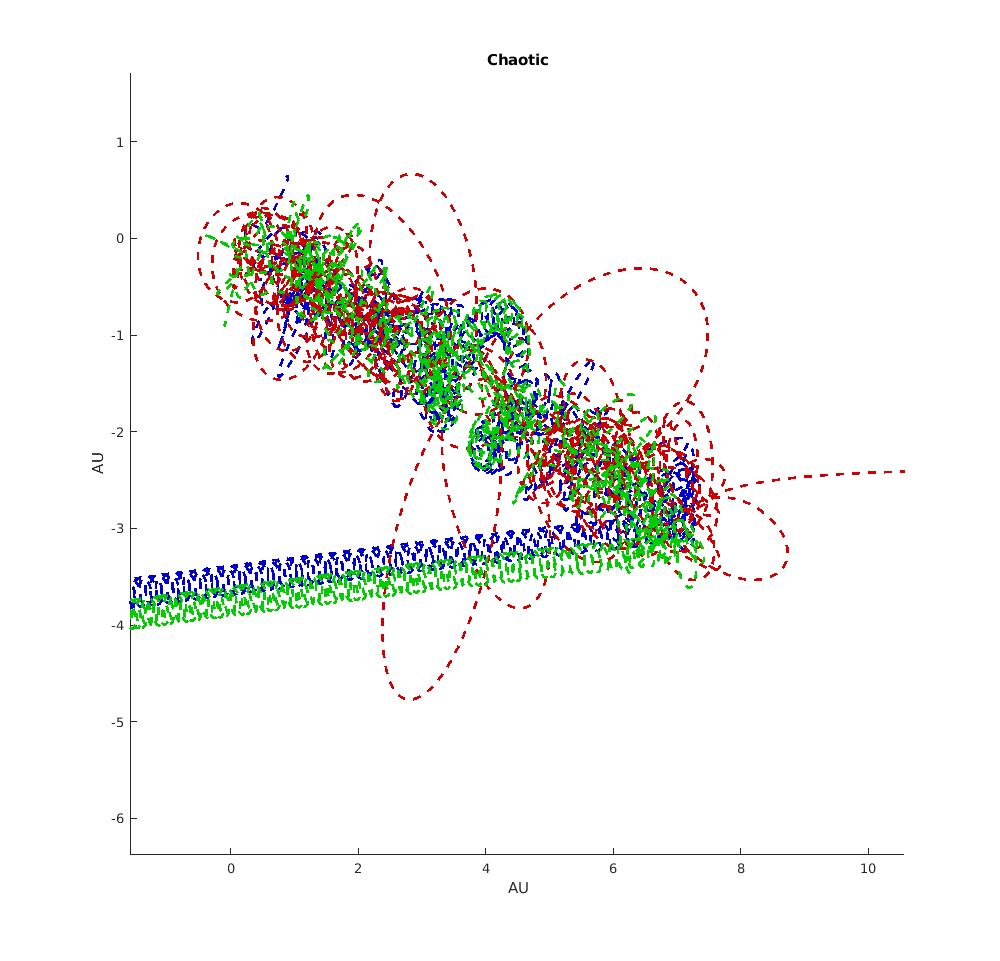} 
        \caption{Chaotic behaviour of an unequal mass system. The highlighting feature displayed in both systems is how the three body system reduces to a two body system to attain stability. The third body is always flung away with the help of the gravity assist of other two bodies.}
    \end{minipage}
\end{figure*}

The systems corresponding to this classification are extremely chaotic in nature. Even the slightest change in the initial conditions ensures that the system traverses a path in the phase space that can never be replicated. Such systems are extremely susceptible to perturbations. 

The model shown below (FIG. 10) demonstrates one of the defining features of chaotic three body systems, the reduction of the complexity of the system. Every system in the universe tries to attain stability and these systems are no different. They obtain stability by reducing from a three body system to a two body system by flinging away the third body.

Another system with completely different initial conditions, unequal masses and a planar restriction is shown below (FIG. 11). The outcome remains the same, in the sense that it's impossible to reproduce this system without taking into consideration every digit of the mantissa into subsequent calculations.

\section{Conclusion}
Re-coding the Mandelbrot set to form the logistic map provided us with an explanation for why the bifurcations lined up with the real line and probed us to further study it's fractal nature. But a more interesting thing we found about the logistic map was it's chaotic behaviour at certain values of r and the periodically distributed islands of stability, this lead us to probe deeper into chaos theory and the problems it posed. Lorenz demonstrated the onset of chaos and sensitive dependence on initial conditions in his experiment - the Lorenz water wheel. One of the widely studied application of chaos is the three body system and the N body system as a whole.

The three body problem lacked an analytical solution for centuries and Poincaré proved why it won't ever be solved analytically. In 1912, Karl Sundman found a series solution\cite{sundman} for the problem but it converges so slowly that it is infeasible to produce practically relevant results from it. Since then the three body problem has been solved for several restrictive cases and the search is still going on for more viable and general systems. Classifying three body systems into long term stable, quasi-stable and chaotic systems helps to not only bring to light the difficulty in finding solutions for such systems but also it's complexity and the infinite solutions it has in certain regions of initial conditions. These can be equated to the islands of stability, only the problem here is that these islands are extremely scattered and thus give us a very small window of exploring them.

\section{Acknowledgements}
We would like to thank our mentor Andres Lopez Moreno for guiding and supporting us throughout the project. We would also like to thank ISEC for providing us this opportunity to be able to research and review topics on our own under expert guidance. I would also like to thank my colleague Shobhit Ranjan for proofreading the article.

\bibliographystyle{apsrev}
\bibliography{main}

\end{document}